\documentclass[12pt]{iopart}
\usepackage{graphicx}

\begin{document}
\sloppy

\title{The Population of Deformed Bands in $^{48}$Cr by emission of
  $^{8}$Be from the $^{32}$S~+~$^{24}$Mg reaction}
\author{S. Thummerer$^1$, 
W.~von~Oertzen$^{1,2}$,
B.~Gebauer$^1$,
S.M.~Lenzi$^3$,
A.~Gadea$^4$,
D.R.~Napoli$^4$,
C.~ Beck$^5$,
M.~Rousseau$^5$}

\address{$^1$ Hahn-Meitner-Institut Berlin, Glienicker Strasse 100,
  14109 Berlin, Germany}
\address{$^2$ Freie Universit\"at Berlin, Fachbereich Physik,
  Arnimallee 14, 14195 Berlin, Germany}
\address{$^3$ INFN, Laboratori Nazionali di Legnaro, Via Romea 4,
  35020 Legnaro(PD), Italy}
\address{$^4$ Dipartimento di Fisica and INFN, Padova, Italy}
\address{$^5$ Institut de Recherches Subatomiques,
  IN2P3-CNRS/Universit\'e Louis Pasteur, B.P. 28, F-67037 Stasbourg
  CEDEX 2, France}

\begin{abstract}
  Using particle~-~$\gamma$ coincidences we have studied the
  population of final states after the emission of
  2~$\alpha$-particles and of $^{8}$Be in nuclei formed in
  $^{32}$S+$^{24}$Mg reactions at an energy of $\textrm{E}_{\rm
    L}(^{32}\textrm{S}) = 130\,{\rm MeV}$. The data were obtained in a
  setup consisting of the GASP $\gamma$-ray detection array and the
  multidetector array ISIS. Particle identification is obtained from
  the $\Delta$E and E signals of the ISIS silicon detector telescopes,
  the $^{8}$Be being identified by the instantaneous ``pile up'' of
  the $\Delta$E and E pulses. $\gamma$-ray decays of the $^{48}$Cr
  nucleus are identified with coincidences set on 2~$\alpha$-particles
  and on $^{8}$Be. Some transitions of the side-band with
  $K^\pi=4^{-}$ show stronger population for $^{8}$Be emission
  relative to that of 2~$\alpha$-particles (by a factor $1.5-1.8$).
  This observation is interpreted as due to an enhanced emission of
  $^{8}$Be into a more deformed nucleus. Calculations based on the
  extended Hauser-Feshbach compound decay formalism confirm this
  observation quantitatively.
\end{abstract}



\section{Introduction}\label{sect:intro}

Recent studies of particle~-~$\gamma$ coincidences using large $\gamma$-ray
detector arrays in conjunction with 4$\pi$ solid state particle-detector balls
have shown that the particle trigger can be a decisive tool in identifying the
final nucleus for a detailed nuclear spectroscopy
\cite{lenzi96,cameron96,cameron98,svenson98}. The enhanced emission of charged
particles from the compound nucleus (CN) has also been discussed as
a possible trigger for more strongly deformed shapes of the compound nucleus
\cite{blann80,blann81,jaenk99}. For lighter CN, in particular for N=Z
$\alpha$-cluster nuclei, the emission of heavier fragments has been observed,
and the process of the emission of $^{8}$Be and  $^{12}$C nuclei is expected to
be strongly enhanced if the CN is well deformed
\cite{sanders87,sanders89,sanders94}. The emission of these ``light''
fragments (also called intermediate mass fragments) is actually viewed in close
relation to asymmetric fission, which may proceed through strongly
deformed shapes \cite{sanders99,nouicer99}. 

\begin{figure}[ht]
  \begin{center}
    \includegraphics[width=.6\textwidth]{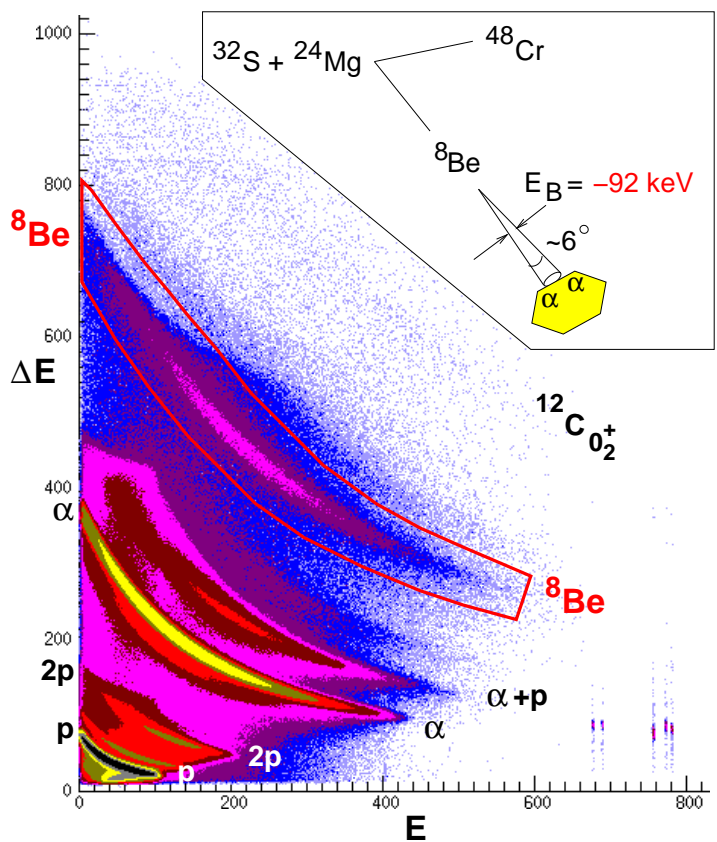}
    \caption{Plot of the $\Delta$E-E signals from the ISIS telescopes
      as obtained from the experiment with $^{32}$S~+~$^{24}$Mg at
      E$_{\textrm{L}}$ = 130~MeV. Events for the hydrogen isotopes p,
      d and t as well as for $\alpha$-particles can be identified. The
      $^{8}$Be events are seen at twice the amplitudes of the
      $\alpha$-particles. The inlet shows the breakup of
      $^8\textrm{Be}_{g.s}$ into 2~$\alpha$-particles and their
      detection in the same detector (see text).}
    \label{fig:dee}
  \end{center}
\end{figure}

We have started a study of the population of highly deformed bands in
$\alpha$-cluster nuclei by using binary reaction
triggers~\cite{gebauer98,thummerer99}. We have chosen reactions
between light $\alpha$-cluster nuclei populating CN in the A=40-56
range, for which very deformed shapes are predicted
\cite{leander75}. In this paper we report results on the emission of
$^{8}$Be in the $^{32}$S~+~$^{24}$Mg reaction, which can be viewed as
a binary process, at $\textrm{E}_{\rm L}(^{32}\textrm{S}) =
130\,{\rm MeV}$, the $^{56}$Ni CN being populated at an excitation
energy of E$^{*}$ = 70~MeV.
The $^{56}$Ni nucleus is of special interest in the sense that it can
be populated by the mass-symmetric reaction $^{28}$Si+$^{28}$Si, which
is known to be one of the most favorable nuclear systems for the
observation of resonant behavior~\cite{betts81c}.  Futhermore, the
$^{28}$Si+$^{28}$Si scattering has revealed triaxial quasi-molecular
states~\cite{nouicer99,beck01} populated by the conjectured $J^\pi =
38^+$ resonance~\cite{betts81c}.
The symmetric and near-symmetric fission processes induced by the
$^{32}$S~+~$^{24}$Mg reaction have been already studied extensively in
the recent past
\cite{sanders87,sanders89,sanders94,sanders99}. Previous $\gamma$-ray
spectroscopy studies of this reaction, using the $\gamma$-spectrometer
GASP (with an photopeak efficiency $P_{Ph}=3\%$)
with the light-charged-particle ball ISIS, have furnished
detailed information on the structure of $^{48}$Cr and other
neighbouring nuclei, which were selected by gating on
$\alpha$-particles and protons \cite{lenzi96,cameron98}. We have used
the same original data \cite{lenzi96} to search for
$\gamma$-coincidences with $^{8}$Be.
The $^{8}$Be nucleus is identified by selecting the corresponding
events (as explained in Sect.~\ref{sect:exp_proc}) from the
$\Delta$E-E matrix shown in Fig.~\ref{fig:dee}.

\section{Experimental Procedures}\label{sect:exp_proc}

The decay of $^{8}$Be in its ground state, which is unbound by
0.0919~MeV, gives rise to 2~$\alpha$-particles emitted in the
laboratory system in a narrow cone of less than 5-7~degrees, which are
registered simultaneously in one of the 40~$\Delta$E-E telescopes of
the ISIS multidetector array. The telescopes have an angular opening
of $\Delta\theta\approx 29\;\textrm{degrees}$, each with an
insensitive gap of 6-7~degrees between two single detectors due to the
width of the detector frames
and cover 64\% of $4\pi$, whereas for the beam entrance and exit each
one detector telescope has been taken out leading to an angular range
from $\theta \simeq 16-164\;\textrm{deg}$.
The $^{8}$Be events are thus easily registered as ``pile up'' events
of 2~$\alpha$-particles.
The Li-isotopes could not be detected due to absorbers ($12\mu m$ Al foil)
placed in front of the ISIS detectors, which modify also the
$\Delta$E-E curves of the detected particles.
The detection efficiency for $\alpha$-particles at the forward 
rings, which carries almost all of the statistics, is not affected
by this absorbers.

The first excited state of $^{8}$Be at an excitation energy of
3.04~MeV, having a width of 1.5~MeV, will produce an emission cone of
more than 30~degrees. This means that only a small fraction of the
$2\alpha$-pairs are registered in one ISIS telescope ($\approx 50\%$).
A possibility to select such events is the selection of
two~$\alpha$-particles in two adjacent telescopes, a procedure, which
has also been pursued. These events show no difference in the
$\gamma$-spectrum to the randomly distributed $\alpha$-particle
triggers, in contrast to those obtained with $^8$Be-trigger as shown
in Fig.~\ref{fig:single_gamma}.  However, due to the high fraction of
chance coincidences in the case of using two adjacent telescopes (see
discussion below), it is necessary to ask e.g. for the sum of the
kinetic energies of both registered $\alpha$-particles in order to
reduce the background of chance coincidences, but the two-dimensional
($E_{\alpha_1}/E_{\alpha_2}$) distribution shows no distinct structure
for $^8$Be$_{0^+}$ or $^8$Be$_{2^+}$ states, and the energy resolution
of the detectors (due to the large opening angle of the single
telescopes, resulting in the coverage of a wide range of angles and
$\alpha$-energies) did not allow to make a selection of $^8$Be states.
Therefore these events of 2~$\alpha$-particles in two adjacent
detectors were not included in the data analysis.  In the later
analysis of the $\gamma$-spectra triggered with $^8$Be a
$^8$Be$_{2^+}$ contribution can not be excluded, but it is not
considered in the compound nucleus calculations in
Sect.~\ref{sect:ehfm}.

An important aspect of the identification of $^8$Be is the
determination of the background of chance coincidences resulting from
2~$\alpha$-particles recorded in the same event in one of the N=40
ISIS telescopes in the $\Delta$E-E branch for $^8$Be. These
2~$\alpha$-particles can either be emitted, (i) from the same event
or, (ii) if emitted from two subsequent different events each emitting
one $\alpha$-particle, they must fall into the rise time of the
electronics of the detection system of $\Delta t\approx
0.1\,\mu\textrm{s}$ (``pile up'' events).

The rate of chance coincidences of the latter type can be estimated by
the expression $R^{ch}_{2\alpha}=R_\alpha\cdot R_\alpha\cdot\Delta t$,
with $R_\alpha$ being the total rate for detection of single
$\alpha$-particle. From the total ISIS rates
$R_\alpha=488\,\textrm{cps}$ (counts/sec),
$R^{ch}_{2\alpha}=0.024\,\textrm{cps}$ and
$R_{^8\hspace{-0.5mm}Be}=9.95\,\textrm{cps}$ (integrated over all N=40
telescopes) we obtain the small average background value
$R^{ch}_{2\alpha}/R_{^8\hspace{-0.5mm}Be}\simeq 0.24\,\%$, where
$R_{^8\hspace{-0.5mm}Be}$ was not corrected for the background events
included. For the most forward ISIS telescope ring, showing the
highest counting rate (96\% for the $2\alpha$ case), the $\alpha$ and
$^8$Be rates and the background percentage can be given as a average
value for a single detector by $R_\alpha=77.9\,\textrm{cps}$,
$R_{^8\hspace{-0.5mm}Be}=1.59\,\textrm{cps}$ and
$R^{ch}_{2\alpha}/R_{^8\hspace{-0.5mm}Be}\simeq 0.04\,\%$,
respectively.

The chance coincidence background
due to two sequentially emitted $\alpha$-particles from the same event
hitting the same detector
can be estimated
from the multiplicity distribution, the solid angle $\Omega$, and the
efficiency $\varepsilon$ of the individual telescopes. We choose the
experimentally observed $\alpha-\alpha$ coincidences in two different
telescopes as a reference point, which fixes the multiplicity of the
events to $M_{\alpha} = 2$, with some very small contributions from
higher multiplicities. The original average multiplicity is close to
$1.05$.  As a first estimate we assume isotropic emission of both
$\alpha$-particles in the laboratory frame (thus neglecting e.g.
kinematics). The total rate to observe a second $\alpha$ in
coincidence with the first is given by the half of the total 
matrix of $N\cdot N$ combinations minus the diagonal part, which
corresponds to the background in the true $^8$Be events ($N$ is the
number of telescopes). This part of the matrix has $N(N-1)/2 = 780$
elements; the corresponding total number of events measured in the
experiment is $3.0\cdot10^6$ ($2\alpha$) coincidences in 30.7~hours.
The total background consisting of 40 diagonal elements (counters) in
the matrix is determined by the ratio 40/780 and must be scaled with
the number of all observed $1.1\cdot10^6$ counts for $^8$Be. We thus
obtain a value of 14 {\%} of $2\alpha$ coincidences as background, a
result, which is independent of the total efficiency to observe one
$\alpha$-particle.
However, due to the kinematically inverted reaction,
$^{32}$S~$\rightarrow$~$^{24}$Mg, and due to the low energies in the
center of mass system of the $\alpha$-emitting nuclei,
$\alpha$-particle emission is strongly focused into the forward
direction by kinematics.  Another effect to be considered is the fact
that for isotropic emission of the first $\alpha$-particle in the c.m.
system, the emission of the second $\alpha$-particle is expected to be
to some degree correlated in the plane of the emission of the first
$\alpha$-particle. This is reflected in the observation that $2\alpha$
coincidences are stronger in telescopes with opposite directions to
the beam for the most forward ISIS ring, enhanced by kinematical
focusing towards $\theta_{c.m.}=0^\circ$ and $180^\circ$.  The
enhanced counting rates at $\theta_{c.m.}=0^\circ$ and $180^\circ$
could not be analyzed quantitatively, because of the beam exit hole in
the ISIS ball and due to the low energies of the backwards emitted
$\alpha$-particles in the laboratory system, which fall below the
detection threshold. We observe 96\% of the hits within the first ring
of ISIS telescopes and 30\% of the second $\alpha$-particles are
detected by ISIS telescopes on the opposite side in the ball within
the same ring.  From this fact the chance coincidence rate of 2~$\alpha$-particles
in one detector given above with 14\% is estimated to be reduced by a
factor of $\approx 2$.
This factor results from the comparison of the theoretically estimated
probability within the first ring of 6~detectors with a fixed detector
for the first $\alpha$, to find the second $\alpha$ in the same detector,
and the measured value for the probability to find the second $\alpha$ in 
the neighbouring detector. The latter should be very close to the
probability to find it in the same detector (which is not directly
measurable).
The most intense $\alpha-\alpha$ coincidence rate for a given
detector has been found not in the neighboring detector, but in the
opposite side (reduction by typically a factor~2).
The resulting value of 7\% of the chance coincidence contribution of
2~$\alpha$-particles in one detector is close to the measured
value of finding the second $\alpha$ in other neighbouring detectors using
the data of all adjacent detectors (i.e. also of the second ring),
which is found to be 6\%.

\begin{figure}[ht]
  \begin{center}
    \includegraphics[width=.4\textwidth]{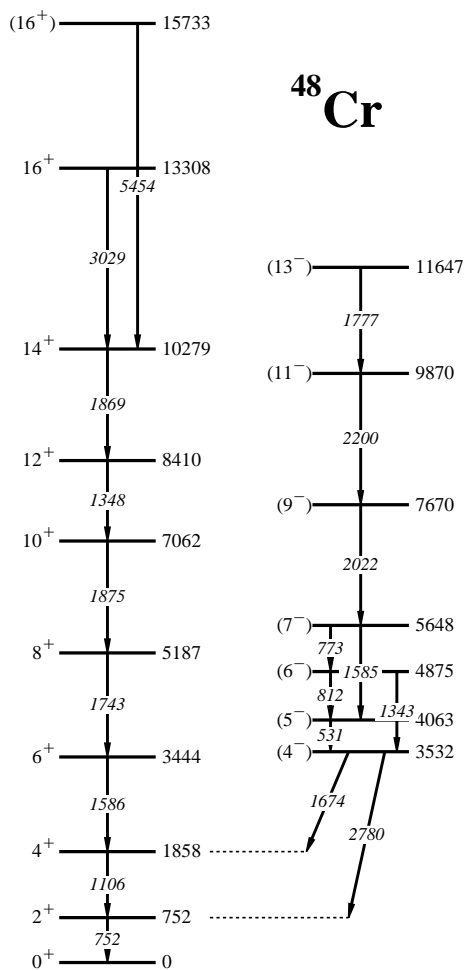}
    \caption{Decay scheme for $^{48}$Cr as obtained from
      Ref.~\cite{brandolini98} with the $\gamma$-ray transitions used
      in the presentation of the relative yields in
      Fig.~\ref{fig:48cr_ratiogs} and \ref{fig:48cr_ratiogssb}, where
      they are labeled by the decay energies shown in this figure.
      This level scheme shows only one excited side band.}
    \label{fig:48cr_scheme}
  \end{center}
\end{figure}
In the following we will use the population of $\gamma$-ray decay
bands in $^{48}$Cr (the decay scheme is shown in
Fig.~\ref{fig:48cr_scheme}) selected by coincidences with the
multidetector array of Compton suppressed Ge $\gamma$-ray detectors,
GASP, for a study of the properties of the binary decay channel,
$^{56}$Ni$\rightarrow$~$^{48}$Cr+$^{8}$Be. We note that also another
binary channel can be selected with the prompt emission of
3~correlated $\alpha$-particles from the second $0^+_2$ in $^{12}$C at
$\textrm{E}^*=7.654\,\textrm{MeV}$, which decays by the emission of
3~$\alpha$-particles (the $\gamma$-ray decay branch is
$8.5\cdot10^{-3}$) with a total energy of 0.196~MeV. They are
registered in one $\Delta$E-E telescope as a triple pile up (see
Fig.~\ref{fig:dee}). $\gamma$-ray spectra of $^{44}$Ti from this
channel have been analyzed as well~\cite{thummerer99}. 
The triple chance coincidence rate for 3~$\alpha$-particles in the $^{12}$C
$\Delta$E-E branch is $0.26\,\%$ assuming an isotropic emission of the
3~$\alpha$-particles.

The experimental details on the GASP and ISIS detector system can be
found in the previous publications~\cite{lenzi96,lenzi98,farnea97}.

\section{Experimental results for $\gamma$-ray yields connected to
  $\alpha$-particle and $^{8}$Be emission}\label{sect:exp_results}

The main aim of the present study is to test the selectivity of
$^{8}$Be emission as a trigger for $\gamma$-ray spectroscopy.
\begin{figure}[ht]
  \begin{center}
    \includegraphics[width=.6\textwidth]{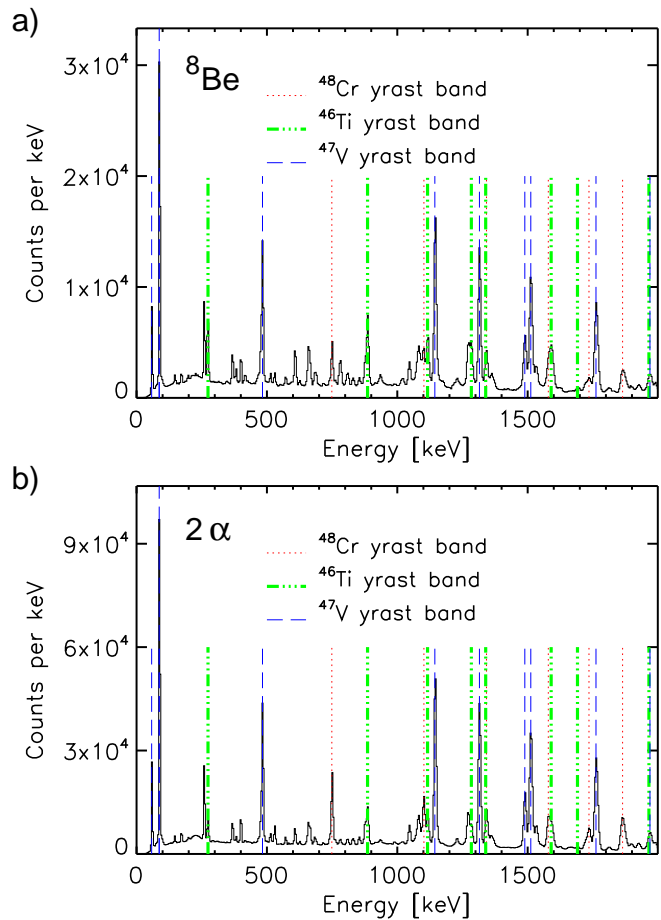}
    \caption{Particle-gated single $\gamma$-ray spectra. (a) Particle
      gate on $^8$Be (b) Gate on 2~$\alpha$-particles. The transissions within 
      the ground state bands of $^{48}$Cr, $^{47}$V and $^{46}$Ti are
      marked.}
    \label{fig:single_gamma}
  \end{center}
\end{figure}
As a first step we look into particle-$\gamma$-coincidence spectra for
2~$\alpha$-particles and $^8$Be.  They are shown in
Fig.~\ref{fig:single_gamma}.  The spectrum is dominated by the
$\gamma$-transitions of $^{47}$V, the spectra are normalized to the
same visual height of the 480~keV doublet line (478~keV and 484~keV)
of $^{47}$V.  We observe very conspicuous differences between the two
spectra: at first sight we recognise that the relative strength of the
three residual nuclei is different: $^{48}$Cr is reduced relative to
$^{47}$V in the case of $^8$Be.  Further $^{46}$Ti is enhanced in the
case of $^8$Be, also relative to $^{47}$V.  The origin of these
differences can be found in the difference in the excitation energy
carried away by 2~$\alpha$-particles as compared to $^8$Be; actually
the compound decay calculations indicate that $^8$Be leaves the
residual nucleus ($^{48}$Cr) at higher excitation energy, thus
partially explaining the enhanced emission of protons from
$^{48}$Cr$^*$ towards $^{47}$V and $^{46}$Ti.  In order to test the
differences in the population in excitation energy and angular
momentum it is useful to have a good know\-ledge of the decay scheme
of the residual nuclei $^{48}$Cr, $^{47}$V and $^{46}$Ti.  In the
further discussion we will use the known level scheme of
$^{48}$Cr~\cite{brandolini98} which is shown in
Fig.~\ref{fig:48cr_scheme}.  Similar analysis has been made for the
cases of $^{47}$V and $^{46}$Ti~\cite{thummerer99}, which are
populated by the additional emission of one or two protons.

\begin{figure}[ht]
  \begin{center}
    \includegraphics[width=.7\textwidth]{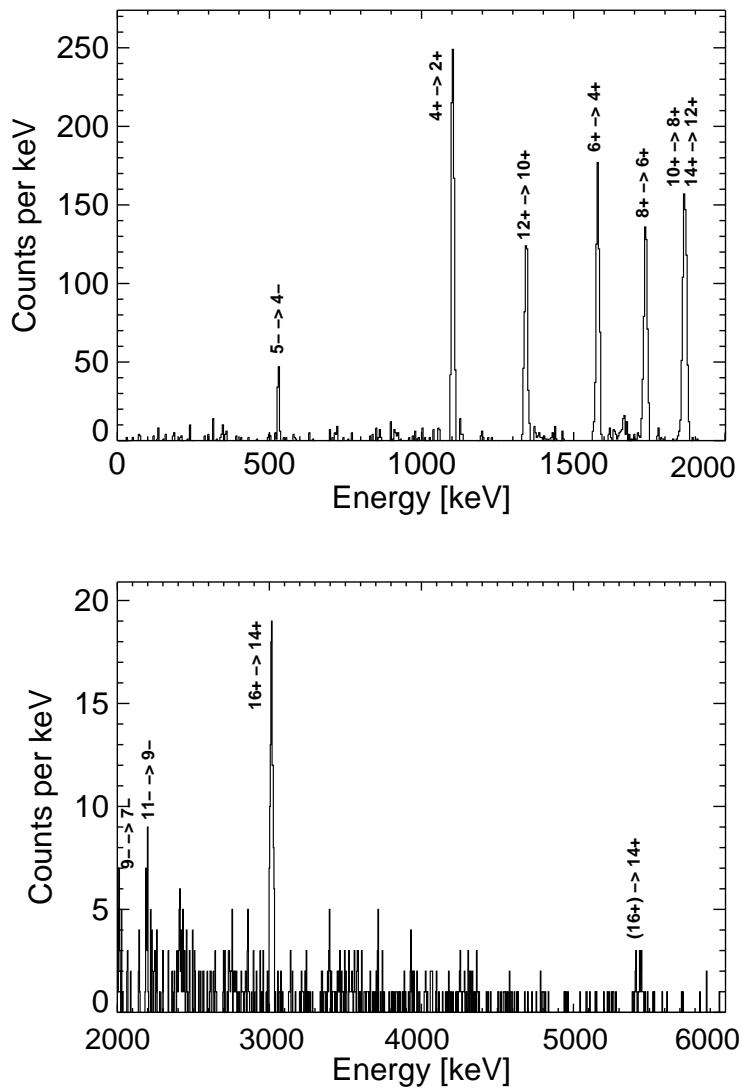}
    \caption{$\gamma$-ray spectra of $^{48}$Cr obtained with a trigger
      on $^{8}$Be in $\Delta$E-E and a gate set on the $2^+\rightarrow
      0^+$ transition showing transitions in the ground state band and
      the $K=4^{-}$ band.}
    \label{fig:spec_cr48}
  \end{center}
\end{figure}

For a discussion of the individual decay schemes
particle-$\gamma$-$\gamma$ coincidences have to be projected.  For a
clear assignment of $\gamma$-ray transition energies such spectra are
shown in Fig.~\ref{fig:spec_cr48}, where we show the $\gamma$-decay
spectra of $^{48}$Cr.
It should be mentioned that due to the
long lifetime of the $4^-$-state ($\tau\sim 4\;\textrm{ns}$) \emph{it is not
possible to use $\gamma$-gated spectra to extract differences} in
the population of the side band for different particle triggers, because
within the lifetime the $\gamma$-emitting nucleus moves several
centimeters and so the Doppler correction fails.
For this reason all $\gamma$-yields given in the following discussion
were extracted using the particle-gated single $\gamma$-ray spectra.
Another reason to go back to the particle-$\gamma$ matrices
(projections of these are the spectra in Fig.~\ref{fig:single_gamma})
is that the statistics in $\gamma$-$\gamma$-yields would be reduced
due to the efficiency of GASP by a factor of $\approx 33$.
The integration of the $\gamma$-peaks can thus be done with smaller
uncertainties. The analysis is done with the same procedure for both
types of particle triggers. However, we have used the $\gamma-\gamma$
coincidences in order to detect cases with energy overlaps in the
selection of the $\gamma$-transitions.

\begin{figure}[ht]
  \begin{center}
    \includegraphics[width=.55\textwidth]{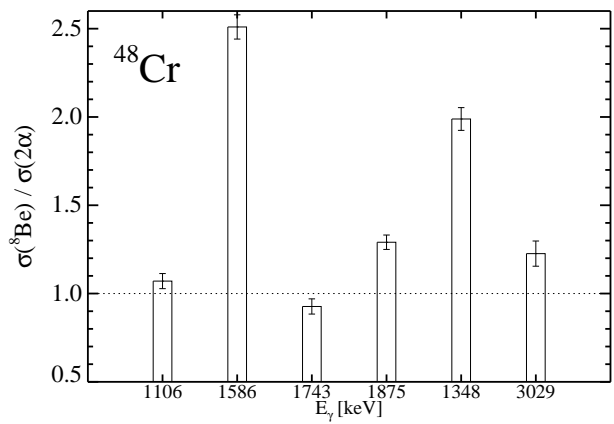}
    \caption{Comparison of experimental $\gamma$-ray yields in the
      gs-band of $^{48}$Cr for triggers on $^8$Be relative to those
      with 2~$\alpha$-particles. The yields have been obtained by
      normalisation to the 752~keV transition (see
      Fig~\ref{fig:48cr_scheme}). $\gamma$-ray transitions are labeled
      by their energies. The error bars contain the statistical errors
      of the used $\gamma$-lines.}
    \label{fig:48cr_ratiogs}
  \end{center}
\end{figure}

For the further analysis the decay scheme of $^{48}$Cr
(Fig.~\ref{fig:48cr_scheme}) is considered.  The ground state band of
$^{48}$Cr is formed by the coupling of the 4~neutrons and 4~protons in
the f$_{7/2}$-shell, respectively, thus the main transitions observed
in the spectrum (Fig.~\ref{fig:spec_cr48}) are connected to the ground
state band reaching up to spin 16$\hbar$, the limit, which can be
deduced from the 8~valence particles outside the N=Z=20 core. A
backbend occurs at a spin value of 10$\hbar$, the calculations of
Caurier et al.~\cite{caurier95} give the correct prediction of this
behavior as well as a value for the quadrupole moment, which points to
the cited deformation. The Nilsson diagram shown in
Fig.~\ref{fig:nilsson} in Sect.~\ref{sect:ehfm}, shows that the mixing
with the d$_{3/2}^+$~shell is at the origin of the deformed shell and
of the rotational character of the yrast levels of $^{48}$Cr up to
spin 16$\hbar$. Inspecting the diagram of Fig.~\ref{fig:nilsson} we
have also the possibility of obtaining a negative parity band by the
excitation of one particle
from the $[202]K=3/2$ to the $[312]K=5/2$
and possibly
from the $[303]K=7/2^-$~Nilsson orbit to the
$[440]K=1/2^+$~orbit of the g$_{9/2}$ shell
for higher deformation. 
The $K=4^-$ band is
expected to have a larger deformation than the ground state band,
because of the gain in energy with increasing deformation.
The first step in the analysis is the consideration of differences in
the population of the higher spins in the ground state band of
$^{48}$Cr. For this purpose the $\gamma$-lines of the ground state
band transitions were fitted in the particle-gated single
$\gamma$-spectra.  The fitting procedure in the high statistics
spectra can be made identical so that differences observed should be
due to differences in the feeding pattern.  The comparison of the this
obtained $\gamma$-yields for the particle trigger on $^8$Be-particles
relative to 2~$\alpha$-particles, obtained by normalisation to the
lowest transition ($2^+\rightarrow 0^+$, 752~keV), is shown in
Fig.~\ref{fig:48cr_ratiogs}.
  The normalization on the $2^+\rightarrow 0^+$ transitions is a
  normalization to the g.s. band: the feeding of the $2^+$ state from
  the $4^-$-state ($\tau\sim 4\;\textrm{ns}$) of the side band (direct
  or via the $4^+$ state) is suppressed, because the nucleus moves
  several centimeters during its lifetime, and therefore the Doppler
  correction causes an additional energy shift which is not taken into
  account in the determination of the peak area of the $2^+\rightarrow
  0^+$ transition, or these $\gamma$'s are not detected due to
  effective opening angle of the Ge's.  In case of a thick target
  experiment, where the $\gamma$-emitting nucleus is stopped, the relative
  normalization as used in this work, would not be meaningful, because
  in that case the $2^+$ state is fed by both bands and the effect of
  an enhancement of the population of the side band would not be seen.

Differences in the yields between the
$^8$Be- and the $2~\alpha$-particle triggers can be seen for the
transitions at 1586~keV, 1875~keV, 1348~keV and 3029~keV. We observe
an enhancement of $25\%-30\%$ for the yields of the case with 1875~keV
($10^+\rightarrow 8^+$, $14^+\rightarrow 12^+$; this is a ``double''
line) and the case with 3028~keV ($16^+\rightarrow 14^+$) transitions
for the $^8$Be-trigger. They are well separated and undisturbed, the
result can be directly interpreted as due to an enhanced population of
higher spins in the case of the emission of $^8$Be from the compound
nucleus.
\begin{figure}[ht]
  \begin{center}
    \includegraphics[width=.55\textwidth]{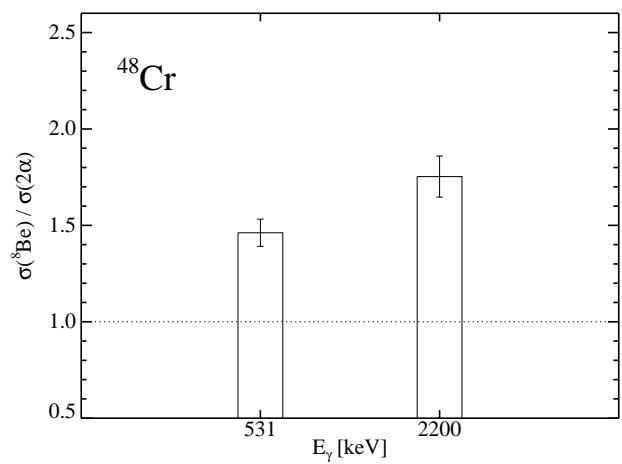}
  \end{center}
    \caption{Experimental ratios of $\gamma$-ray yields in the
      side-band of $^{48}$Cr (obtained as in
      Fig.~\ref{fig:48cr_ratiogs}) for triggers on $^{8}$Be relative
      to those with 2~$\alpha$-particles for $\gamma$-ray transitions
      in $^{48}$Cr labeled by their energies (shown in
      Fig.~\ref{fig:48cr_scheme}).}
    \label{fig:48cr_ratiogssb}
\end{figure}

The enhanced $\gamma$-yields for the 1586~keV and 1348~keV transitions
are influenced by overlapping $\gamma$-lines from another nucleus; the
main contribution of this disturbance is due to unseparated
$\gamma$-lines from $^{46}$Ti at similar energies (1598~keV,
1345~keV), reflecting the stronger population of $^{46}$Ti in the
$^8$Be case (see Fig.~\ref{fig:single_gamma}); only a small
contribution is caused by the existence of side band transitions at
this energies.  The other relative $\gamma$-yields shown in
Fig.~\ref{fig:48cr_ratiogs} (1106~keV and 1743~keV) between low lying
spins in $^{48}$Cr differ only by less than 10\% for the different
particle triggers, i.e. no enhancement appears (as expected) for these
cases.

In the second step of the analysis we considered the differences in
the population of the side band for the different particle triggers.
Fig.~\ref{fig:48cr_ratiogssb} shows yields of $\gamma$-lines of side
band transitions for particle triggers on $^8$Be relative to those
with $2~\alpha$-particle, again each normalised to the $2^+\rightarrow
0^+$-transition (752~keV).  For the side band the $5^-\rightarrow
4^-$- (531~keV) and the $11^-\rightarrow 9^-$ transition (2200~keV)
were used which are free from contaminant lines.  We see an
enhancement by a factor of 1.5 and 1.8, respectively, for the particle
trigger on $^8$Be.

We will later relate this effect to a possible larger deformation of
the side band. Similar observation of an enhanced population of a side
band in a binary decay has been observed in a binary reaction
$^{32}$S+$^{24}$Mg populating $^{28}$Si ($K=0_3^+$-band)~\cite{sanders94}.

It should be mentioned that due to energy resolution and statistical
reasons not all lines of the investigated bands could be used for the
analysis.
Further experiments with higher statistics are necessary to establish
more cases.

We conclude that after the emission of $^8$Be the population in
excitation energy and angular momentum favors
the $K^\pi=(4^{-})$ side band of $^{48}$Cr. The enhanced emission of
$^8$Be into deformed states of the daughter nucleus in statistical
compound decay can be related to the differences in phase space in the
emission process or to a structural preference for the emission of
clusters as it is suggested by cluster model
considerations~\cite{zhang94}. As a first step we will consider the
differences in $2\alpha$ and $^8$Be emission in the framework of the
compound statistical decay using the extended Hauser-Feshbach
method~\cite{matsuse97} described in the next section.

Similar analysis has been made for the neighbouring $^{47}$V
nucleus, which is reached from $^{48}$Cr with high probability by a
subsequent proton emission and also here an enhanced population of the 
deformed side band has been found in the binary case~\cite{thummerer99}.
Since the decay scheme (c.f.~\cite{cameron98}) is more complex than
for $^{48}$Cr and may not be complete, this case is not so well suited 
for the present discussion.

  At the present stage the statistics is still too poor to examine the
  detailed population patterns for the $^{44}$Ti nucleus as populated
  by the $3\alpha$ state of $^{12}$C$_{0^+_2}$ from the
  $^{44}$Ti+$^{12}$C exit channel. Additional experiments are
  currently underway with GASP and EUROBALL to attack this problem.

\section{Extended Hauser-Feshbach calculations for CN decay
  (fusion-evaporation and fusion-fission)}\label{sect:ehfm}
 
In the following the emission of $^{8}$Be from a light CN is
considered in the framework of the Extended Hauser-Feshbach Method
(EHFM) as developed by Matsuse and collaborators~\cite{matsuse97}. In
this approach the binary decay (i.e. fusion-evaporation of
intermediate mass fragments and/or fusion-fission) of the CN is
treated in addition to the well known process of fusion-evaporation of
light particles (neutrons, protons, $\alpha$). The method has been
applied to several light-mass systems~\cite{matsuse97,beck96a,beck98}
in the compound nuclear masses region of A=40-80, and excellent
agreement has been achieved with a standard set of parameters. For
more details on the calculational procedures and a discussion of the
sensitivity to the parameters of the model, we refer to
Ref.~\cite{matsuse97}. However, for a better understanding of the
dependence of the heavy fragment emission some parts of the
formalism, which determine the decay probability, are also described
hereafter. The partial width $\Gamma_J$ for a decay channel with spin 
$J$ is given by the relation
\begin{equation}
  \label{eq:gam}
  \Gamma_{\rm J} = \frac{{P_{\rm J}}}{2\pi\rho_{\rm J}(\varepsilon)},
\end{equation}
where $\rho_J$ is the level density as obtained by the usual Fermi gas
expressions~\cite{bohr81}, $\varepsilon$ the excitation energy in the
daughter nucleus and $P_{J}$ the phase space integral for the decay
into the daughter nucleus. In the latter the transmission coefficients
in the binary channel are of paramount importance, and thus indirectly
the deformation of the two fragments, which will determine the Coulomb
barrier. The total potential between the two fragments is parametrized
as
\begin{equation}
  \label{eq:pot}
  V(L)  =  V_{\rm Coul} + \frac{\hbar^2}{2\mu_{\rm f} R^2_{\rm S}}L(L+1)
\end{equation}
Here the following definitions are used: $\mu_{\rm f}$ -- reduced mass
in the center of mass (cm) system, $R_{\rm S}$ -- the distance between
the two fragments, which is defined as consisting of the two radii
$R_{\rm L}$ and $R_{\rm H}$ and a neck-distance parameter $d$ given by
\begin{equation}
  \label{eq:Rs}
  R_{\rm S} =  R_{\rm L} +  R_{\rm H}  + d 
\end{equation}

  From a recent systematic study~\cite{nouicer97,beck99} of a large
  number of heavy-ion fusion-fission reactions (i.e.
  $^{28}$Si+$^{12}$C~\cite{rousseau01,beck00},
  $^{35}$Cl+$^{12}$C~\cite{beck96a},
  $^{36}$Ar+$^{12}$C~\cite{farrar96},
  $^{28}Si$+$^{28}$Si~\cite{rousseau01},
  $^{32}$S+$^{24}$Mg~\cite{sanders89} and
  $^{35}$Cl+$^{24}$Mg~\cite{beck98}) the given initial value of $d$ used
  here to be $d=3.0\pm 0.5\;\textrm{fm}$, has been found
  to have the following linear dependence with CN mass~\cite{beck99}:
  $d=0.112(A_{CN}-24.65)$.
For $R_{\rm L}$ and
$R_{\rm H}$ the radius parameter $r_0 = 1.2\,\textrm{fm}$ is commonly
used ($R_{\rm L,H}=r_0\,A_{\rm L,H}^{1/3}$) \cite{matsuse97,beck99}.
In order to explain the present data consistently the adopted
mass-dependence of the neck-distance parameter d, which is the only
adjustable parameter of the EHFM scission point picture, is introduced
to simulate the variation of the CN deformation. It has been
shown~\cite{sanders99} that the shapes of the saddle- or
scission-points, as assumed with the transition-state model and EHFM,
respectively, are almost identical.
A longer neck, corresponding to a
larger deformation, will give a reduced barrier height and thus a
larger transmission probability. The deformation of the decaying
system will thus enhance the emission of heavy fragments as already
suggested by Blann~\cite{blann81}.

The second effect, which will have a significant influence on the phase space
integral is the level density $\rho_{J}(\varepsilon)$ in the daughter nucleus
at spin $J$ and excitation energy $\varepsilon$, 
\begin{equation}
  \label{eq:rho}
  \rho_{J}(\varepsilon) = \frac{1}{12} \left(\frac{a\hbar}{2\theta}\right)^{3/2}(2J+1)a
  \frac{e^{2\sqrt{X}}}{X^2}
\end{equation}
\begin{equation}
  \label{eq:ix}
   X = a\left(\varepsilon-\frac{\hbar^2}{2\theta}J(J+1)-\Delta_{\rm pair}\right)
\end{equation}
The important parameters here are the level density parameter $a$ and
the moment of inertia $\theta$, which determines the position of the
yrast line of the daughter nucleus. For larger deformations the yrast
line bends down, a larger phase space is obtained for the decay.
  In the initial version of EHFM~\cite{matsuse97}, the measured ground
  state binding energies are used to evaluate the excitation energy of
  the decaying fragment, but an average level density parameter was
  introduced. However, two sets of parametrizations are available for
  the level density parameter {\it a} used in each step of EHFM. A
  constant value ({\it a}~=~A/8) was chosen for the preliminary
  calculations of Ref.~\cite{matsuse97}. This parameter set may
  overestimate shell effects. An alternative way to reproduce the
  strong variation from fragment to fragment is to incorporate shell
  effects in the level density formulas. In the present work (as well
  as in Refs.~\cite{beck96a,beck98,nouicer97,beck99}) shell
  corrections in the energy-dependent (temperature-dependent) level
  density parameter {\it a} are produced by the difference of the
  experimental mass and the liquid drop mass for each fragment. In
  order to introduce the shell effects in the level density parameter
  ${a}$, we use an improved version~\cite{beck99} of the empirical
  formula of Bohr and Mottelson~\cite{bohr81}. The formula evaluates
  the shell structure energy which depends on the nuclear temperature
  ${T}$ and consequently the level density parameter ${a}(T)$ becomes
  nuclear temperature dependent. To get the shell structure energy of
  the ground state, we use (see Refs.~\cite{beck98}, for instance) the
  measured ground state binding energies and the liquid drop binding
  energies.  This new modelization is currently being developed and
  will be discussed with the original EHFM
  parametrization~\cite{matsuse97} in more detail in a forthcoming
  publication~\cite{beck99}.

\begin{figure}[htbp]
  \begin{center}
    \includegraphics[width=.7\textwidth]{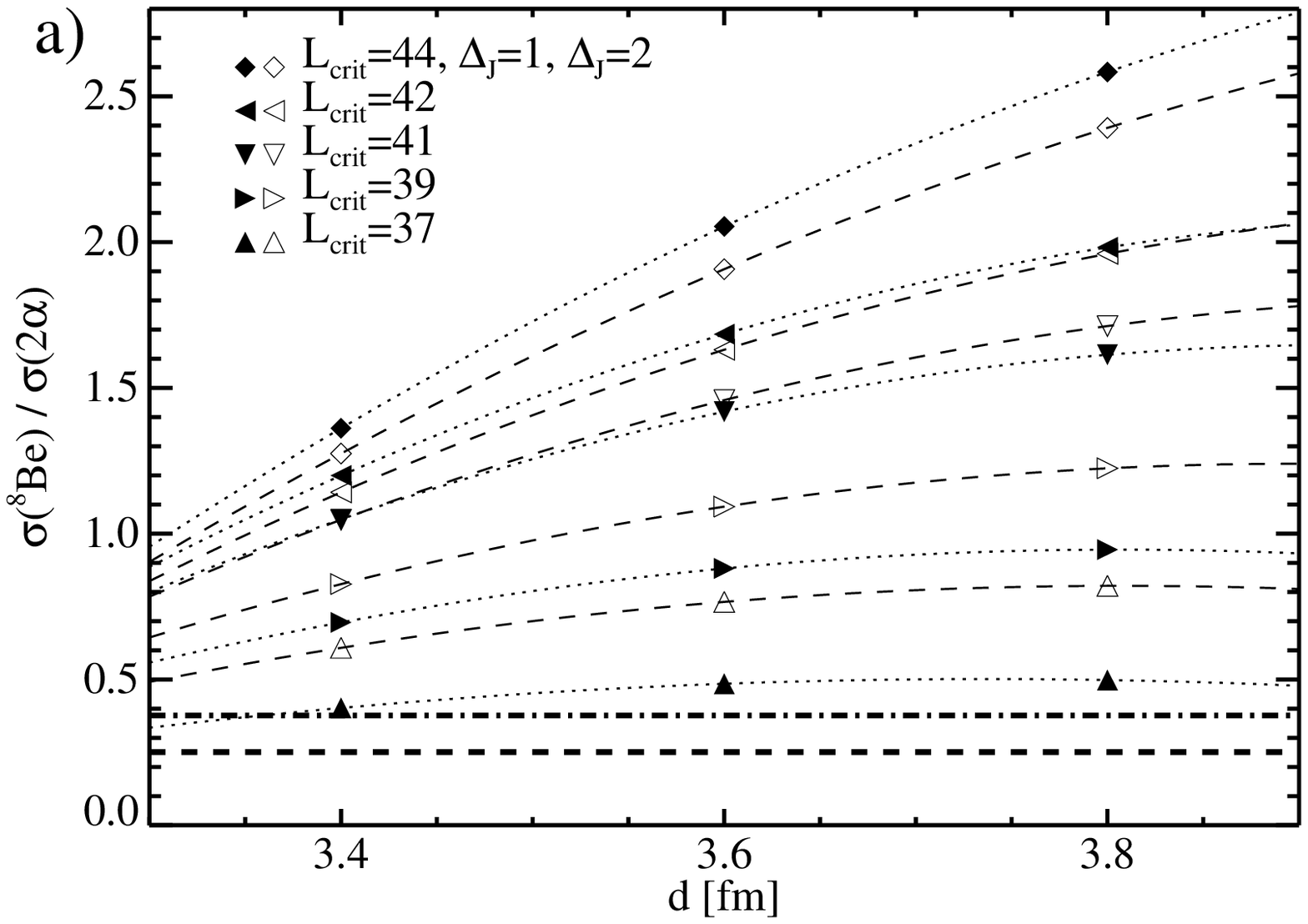}
    \includegraphics[width=.7\textwidth]{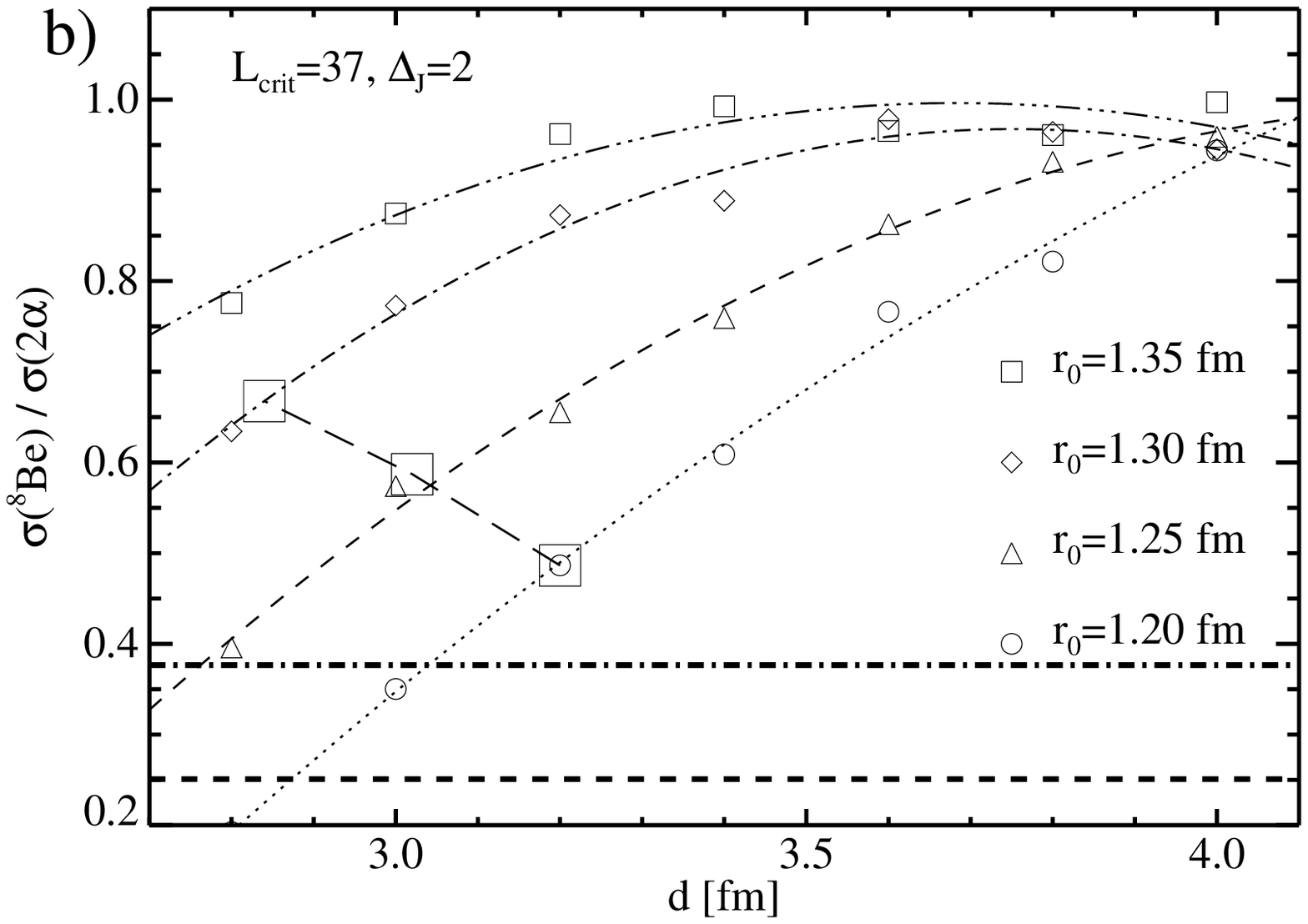}
    \caption{Ratio for the emission of $^{8}$Be relative to two
      uncorrelated $\alpha$-particles as obtained from the EHFM
      calculations as function of the deformation in a binary
      split, parametrized by the parameter $d$ (neck size).
      a) Variation of $L_{cr}$ and $\Delta_J$. The filled symbols
      correspond to $\Delta_J=1$, the open ones to $\Delta_J=2$. The
      radius parameter $r_0$ was set to $1.2\;fm$.
      The experimental value of the ratio for the whole
      data set is indicated by the horizontal line. The second
      horizontal line is given by the enhancement factor determined
      from the $\gamma$-ray spectra as described before.
      b) Variation of $r_0$ for fixed values $L_{cr}=37$ and
      $\Delta_J=2$. The squares mark the $r_0,\,d$ values for constant 
      value of $R_{\rm s}$ (Eq.~\ref{eq:Rs}) but different moments of
      inertia.}
    \label{fig:ehf}
  \end{center}
\end{figure}

For the variations in the transmission coefficients in the binary
channel we have varied the total distance $R_{\rm S}$ using the neck
size parameter $d$, and the
radius parameter $r_0$. The results are illustrated in
Fig.~\ref{fig:ehf}. We show actually the relative yield of $^{8}$Be to
2~$\alpha$-particles, both popula\-ting the same final $^{48}$Cr
nucleus. The dependence on other parameters of the EHFM calculations
are also shown: the primary L-value distribution determined by the
maximum value of $L$, $L_{\rm cr}$, and by its diffuseness $\Delta_J$.

The experimentally determined ratios $\sigma
(^{8}\textrm{Be})/\sigma(2\alpha)$ are shown by the horizontal lines
in Fig.~\ref{fig:ehf}.  The dashed line indicates the experimental
value of the ratio for the whole data set, the dashed-dotted line is
given by the enhancement factor determined from the $\gamma$-ray
spectra as described before.  It is difficult to compare this value
directly with the calculated value, because the experiment does not
cover the full $4\pi$ range (as described in
Sect.~\ref{sect:exp_proc}), which is obtained in the calculation.  In
order to compare with the experiment we have to discuss the
\emph{relative changes} of the calculated ratio induced by the
variation of the neck size parameter $d$, or by modifications in the
angular momentum distribution in the incident channel. The result is
shown in Fig.~\ref{fig:ehf}a. It can be observed that for the maximum
spin values for the formation of the $^{56}$Ni CN,
$L_{cr}=37-39\hbar$, adopted from Ref.\cite{matsuse97} (consistent
with the available measured complete fusion cross
sections~\cite{hinnefeld87}), the $^8$Be emission is not favoured
relative to the $2\alpha$ emission for more deformed states (at
$d\approx3.8\;\textrm{fm}$). This result has also been observed in the
work of Blann and Komoto \cite{blann81}.  Actually $^8$Be carries more
angular momentum and Fig.~\ref{fig:ehf}a shows that its emission will
be favored at higher angular momenta ($L_{cr}>40\hbar$) for larger
deformations, as also predicted in Ref.~\cite{blann81}. It is
interesting to note that the limiting angular momentum for $^{56}$Ni
is experimentally found to be close to 42$\hbar$~\cite{beck96b} in
agreement with the predictions of a modified version of the rotating
liquid drop model (LDM), which includes finite-range corrections of
the nuclear interaction by means of a Yukawa-plus-exponential
attractive potential and diffuse-surface effects~\cite{sierk86}.

We consider
in addition
then a second effect, which is also discussed in
Ref.~\cite{blann81}, the increased phase space for
the
more strongly
deformed daughter
nuclei,
i.e. for a softer yrast line due to a
larger moment of inertia. In the present formulation this can be
achieved by an increase of the radius parameter $r_0$ of the heavy
fragment $R_{\rm H}$. The result for different values of $r_0$ is
shown in Fig.~\ref{fig:ehf}b, again as function of $d$, the neck
parameter. The figure shows appreciable effects for changes of $r_0$.
We can read from this graphs also the effect produced by the larger
moment of inertia alone by choosing $R_{\rm S}(d,r_0)=const$. Thus the
constant values of $R_{\rm S}$ for
($d=3.2\;\textrm{fm},\;r_0=1.20\;\textrm{fm}$), for
($d=3.02\;\textrm{fm},\;r_0=1.25\;\textrm{fm}$) and for
($d=2.84\;\textrm{fm},\;r_0=1.30\;\textrm{fm}$) must be considered
(they are marked as square symbols in Fig.~\ref{fig:ehf}b) and we
deduce an enhancement of the $^8\textrm{Be} / 2\alpha$ ratio by a
factor $\sim1.7$ for an increase of $\Theta$ by the factor $1.3 / 1.2
\sim 1.1$, which implies that for the present situation an increase of
the moment of inertia of 10\% of the daughter nucleus is reflected in
a noticeable enhancement of the $^8$Be yield by a factor 1.7.  From
these calculations we conclude that the changed ratio of the emission
of $^{8}$Be relative to that of two consecutive $\alpha$-particles can
be related to an increase of the deformation parameter $\beta_2$ of
the heavier fragment, for which a discussion is given in the following
section.

\section{Decay scheme and deformed bands in $^{48}$Cr}\label{sect:decay_scheme}

The structure of the $^{48}$Cr nucleus is described quite well by the
consideration of the half filled f$_{7/2}$ shell, which allows various
nucleon excitations into higher orbitals.

The ground state band is suggested to have a deformation of $\beta_2 =
0.28$~\cite{brandolini98} near the ground state and the deformation
parameter reaches $\beta_2 = 0.1$, as the band termination at $I=16^+$
is approached. This decrease is explained by the fact that because of
the relatively small number of nucleons in the 1f$_{7/2}$ shell, it is
not possible to obtain collectivity at high spins. In the case of the
negative parity band, collectivity remains stronger also at higher
spins due to the mixing with the g$_{9/2}$ shell. Even though the
deformation at low spins is almost equal for the gs-band and the side
band, the deformation at higher spins is larger for the side band.
This is the crucial point when comparing the feeding into $^{48}$Cr,
because it occurs not nearby the ground state but at higher spins.

\begin{figure}[ht]
  \begin{center}
    \vspace*{5mm}
    \includegraphics[width=.9\textwidth]{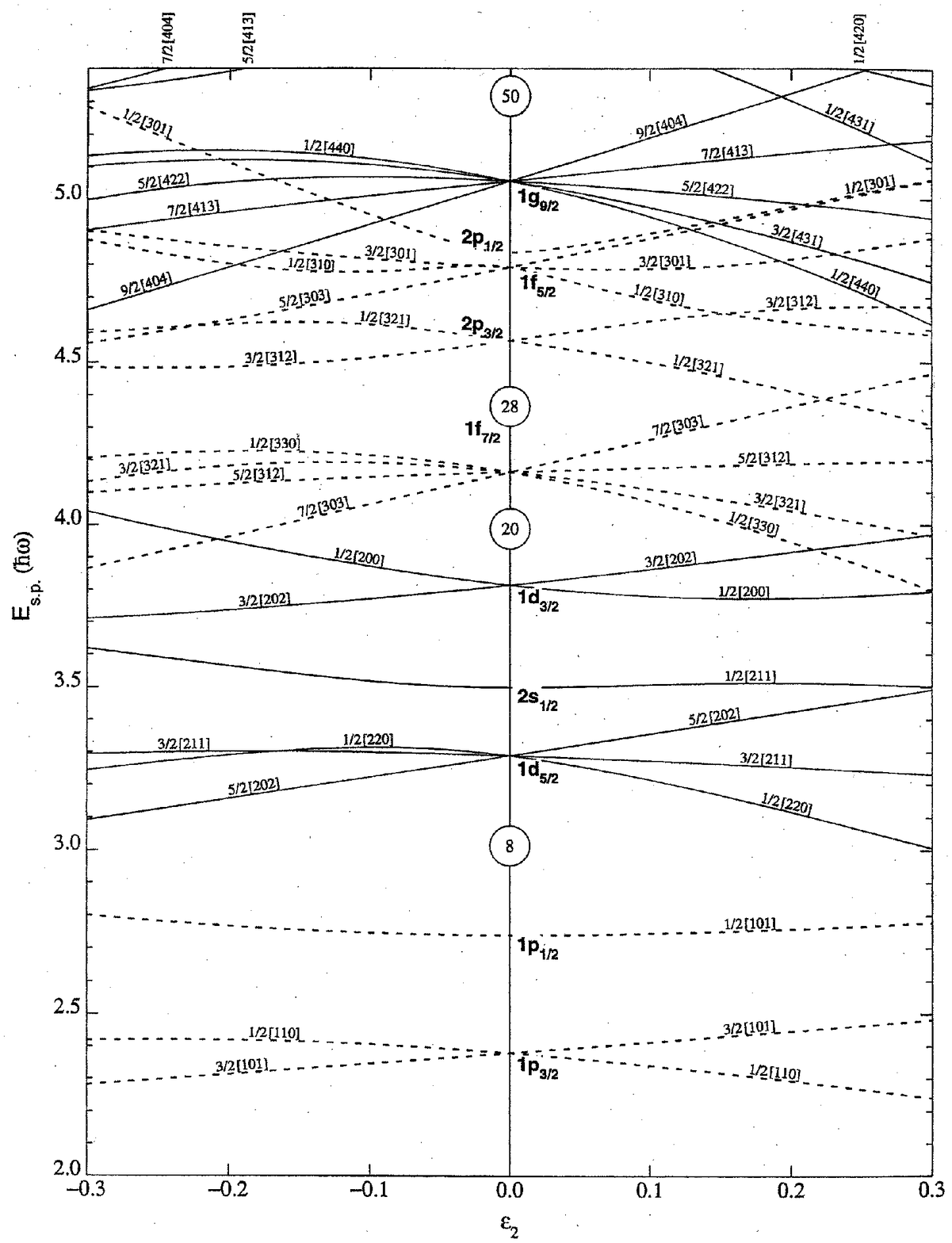}
    \caption{Nilsson diagram, the energies of valence orbitals as
      function of deformation parameter $\beta_2$. Note the small
      distance of the $[303]K=7/2^-$ and the $[440]K=1/2^+$ orbitals
      as the deformation parameter $\beta_2$ ($\epsilon_2$) tends to
      0.35, and the rising energy of the $[202]K=3/2^-$ orbit
      approaching the f$_{7/2}$ shell, which is only half filled for
      $^{48}$Cr.}
    \label{fig:nilsson}
  \end{center}
\end{figure}
The result of the enhanced $\gamma$-ray yields for the $^{8}$Be
trigger for the side-band with negative parity ($K=4^{-}$) can be
related to higher deformations at high spins for this band.
Particle-hole excitations from the d$_{3/2}$ to the f$_{7/2}$ orbit
and also into the g$_{9/2}$ orbit of the next shell will be favored for
larger deformations
and higher spins.
With these configurations a band head of $K=4^{-}$ is obtained. Other
side-bands with positive parity can be expected for the
2-particle--2-hole excitations to the [310]1/2 Nilsson orbital, which
gives a band head with $K=4^{+}$, but higher excitation energy than
for $K=4^{-}$. These features can be read qualitatively from the
Nilsson diagram in Fig.~\ref{fig:nilsson}.

\section{Discussion and conclusions}\label{sect:conclude}

The experimental observation of an enhanced emission of $^{8}$Be into
a more deformed side-band in $^{48}$Cr has been observed by the use of
particle~-~$\gamma$ coincidences using a large $\gamma$-ray detector
array. Using the \emph{Extended-Hauser-Feshbach Method} (EHFM) it has
been shown that cluster emission will be
enhanced at high angular momenta for the decay into deformed residual nuclei.
The result is related to previous theoretical and experimental
studies of cluster emission from compound nuclei. In the work of Blann
and Komoto~\cite{blann81} on cluster emission amplification, actually
the case of the compound nucleus $^{56}$Ni at an excitation energy of 94~MeV,
somewhat higher than ours (70~MeV), has been treated. In their work
deformation is introduced at various stages of the decay (mainly at
the same places as in the present work), and the emission of clusters
up to the mass $^{12}$C is considered. They find that the fractional
decay probability for $\alpha$-particles varies only moderately with
the introduction of deformation (at the maximum spin values of
$40-45\hbar$); at these angular momenta an increase of the emission
probability for $^{8}$Be and $^{12}$C by factors $2-4$ is observed if
the proper deformation values from LDM are introduced. The comparison
of our result
with Ref.~\cite{blann81} has to be seen with the constraint that in the
previous work
finite-range corrections, which produce for each angular momentum a
lowering of the fission barrier due to the attractive forces between
surfaces of the two nascent fragments at the saddle
point~\cite{beck96b}, have not been taken into account. We can
state that the present result and their work~\cite{blann81} are in agreement with
the observation that the emission of heavier fragments at higher
angular momenta is enhanced from and to deformed configurations. Our
calculations have shown that this phenomenon depends on the maximum
spin in the primary population, an effect which is also documented
in~\cite{blann81}.

The main origin of the enhancement in the population of more deformed
shapes after the emission of $^8$Be and $^{12}$C might be linked to
structural effects \cite{freer90}, like specific population of the
states close to the yrast line. This feature will be pursued in a
future investigation, where the population of large deformations of
fragments from a binary process will be searched for in
particle~-~$\gamma$ coincidences with EUROBALL~\cite{gebauer98} and at
higher incident energies.

\ack This work has been started during a sabatical leave of
W.~von~Oertzen at the LNL. He would like to express his gratitude for
the hospitality extended to him during this stay.

\end{document}